# Best Thermoelectric Efficiency of Ever-Explored Materials

Byungki Ryu,[1,*] Jaywan Chung,[1] Masaya Kumagai,[2,3] Tomoya Mato,[4]

Yuki Ando,[4] Sakiko Gunji,[4] Atsumi Tanaka,[4] Dewi Yana,[4]

Masayuki Fujimoto,[4] Yoji Imai,[4] Yukari Katsura,[2,4] and SuDong Park[1]

1, Energy Conversion Research Center, Korea Electrotechnology Research Institute (KERI), Changwon 51543, Republic of Korea

2, Center for Advanced Intelligence Project, RIKEN, Tokyo 103-0027, Japan

3, SAKURA internet Research Center, SAKURA internet Inc., Osaka 530-0001, Japan

4, Research and Services Division of Materials Data and Integrated System (MaDIS), National Institute for Materials Science (NIMS), Ibaraki 305-0044, Japan

*Corresponding Author: byungkiryu@keri.re.kr (B.R.)

Version: arXiv, 20220929

**arXiv:2210.08837v1** [Submitted on 17 Oct 2022]

**arXiv:2210.08837v2** [arXiv number included. Supporting Data files notice included]

**arXiv:2210.08837v3** [effective size of data is updated]

# Abstract

A thermoelectric device is a heat engine that directly converts heat into electricity. Many materials with a high figure of merit *ZT* have been discovered in anticipation of a high thermoelectric efficiency. However, there has been a lack of investigations on efficiency-based material evaluation, and little is known about the achievable limit of thermoelectric efficiency. Here, we report the highest thermoelectric efficiency using 12,645 published materials. The 97,841,810 thermoelectric efficiencies are calculated using 808,610 device configurations under various heat-source temperatures ($T_h$) when the cold-side temperature is 300 K, solving one-dimensional thermoelectric integral equations with temperature-dependent thermoelectric properties. For infinite-cascade devices, a thermoelectric efficiency larger than 33% (≈1/3) is achievable when $T_h$ exceeds 1400 K. For single-stage devices, the best efficiency of 17.1% (≈1/6) is possible when $T_h$ is 860 K. Leg segmentation can overcome this limit, delivering a very high efficiency of 24% (≈1/4) when $T_h$ is 1100 K.

# Main text

**A thermoelectric device** composed of P- and N-type thermoelectric material legs placed between hot- and cold-side substrates can directly convert thermal energy into electrical energy via thermoelectric effects[1]. As the device can generate electric power using low-grade thermal energy[2], thermoelectric technology has been considered particularly promising for industrial or vehicle waste heat recovery systems[3,4].

To put thermoelectric technology into practical use, **it is essential to increase the efficiency ($\eta$) of the devices.** Because a thermoelectric device is a heat engine, the thermoelectric efficiency increases with temperature difference and is bounded by the Carnot efficiency. Moreover, as discovered by Ioffe in 1957, in some cases, the efficiency is determined by a single material parameter[5], called the dimensionless thermoelectric material figure of merit $ZT = \alpha^2 T/\rho\kappa$, which is defined as the ratio between absolute temperature $T$ and three thermoelectric transport properties (TEPs), namely, the Seebeck coefficient $\alpha$, electrical resistivity $\rho$, and thermal conductivity $\kappa$. Precisely, for ideal one-dimensional thermoelectric legs having *temperature-independent* TEPs, the maximum efficiency of thermoelectric conversion ($\eta_{\max}$) under a given operating temperature range from the cold-side temperature $T_c$ to the heat-source temperature $T_h$ is exactly determined by $ZT$:

$$\eta_{\max} = \frac{T_h - T_c}{T_h} \frac{\sqrt{1 + ZT_m} - 1}{\sqrt{1 + ZT_m} + T_c/T_h}, \qquad (1)$$

where $T_m$ is $(T_h + T_c)/2$. From the observation that the higher the $ZT$ is, the higher the efficiency, many high $ZT$ materials have been discovered and developed, from the traditional $Bi_2Te_3$-, $PbTe$-, $GeTe$-, and $Si_{1-x}Ge_x$-based alloys[6–9] to the recently developed SnSe- and $Mg_3Sb_2$-based alloys[10,11].

However, little is known about the current status of device efficiency. The efficiency estimation using $ZT$ is not very accurate for wide-temperature-range applications[12] because TEPs are temperature dependent. Additionally, the estimation ignores several factors introduced in device fabrication processes. Although the measurement of device efficiency is demanded, the number of such experimental studies is less than a few dozen[8], likely because of more complex processes in device fabrication than in materials synthesis. **Therefore, the efficiency-based evaluation of materials and devices is crucial for understanding the current status. Furthermore, determining the achievable limit will guide future directions and accelerate research regarding the design of high-performance devices.**

**Here, we report the theoretical best thermoelectric efficiency of devices resulting from the ever-explored thermoelectric material data collected in the Starrydata2 thermoelectric database.** Starrydata2[13] is the world's largest thermoelectric property database (DB), which contains 43,601 material samples for thermoelectric properties from 7,994 publications as of 2021-November-24. After data filtering, we obtain high-quality thermoelectric *big data* composed of 13,338 samples from 3,120 publications. Then, the material samples in the big data DB are theoretically evaluated by the computed maximum thermoelectric conversion efficiency under various temperature differences of $\Delta T = T_h - T_c$ when $T_c = 300$ K. The thermoelectric efficiency is computed by solving one-dimensional thermoelectric integral equations for the temperature distribution $T(x)$ and heat currents at the hot and cold sides[12,14]. Using the searched high-performance P- and N-leg samples, of which the efficiency is larger than or equal to 85% of the best efficiency, single-stage P-N leg-pair devices with various leg geometries and interfacial resistances are constructed. The interfacial

resistances allow us to include efficiency loss from device fabrication. Finally, the best thermoelectric efficiency for various electrical and thermal operating conditions is theoretically explored over 7,650,225 material efficiency data and 97,841,810 device efficiency data.

## Best thermoelectric efficiency

**Fig. 1 shows the achievable best thermoelectric device efficiency ($\eta^{(\text{dev})}$) among 808,610 P-N leg-pair thermoelectric devices made of 13,353 ever-explored materials,** for a given heat-source temperature $T_h$ and fixed $T_c = 300$ K. The infinite-cascade device, where the electric current at each temperature point is optimized, attains the theoretical maximum efficiency for a given $ZT$ curve. We found the maximally achievable $ZT$ curve ($Z_{\text{max}}(T)$) from the database. Hence, the best efficiency of the infinite-cascade device ($\eta_{\text{max}}^{(\infty-\text{Cascade})}$) is simply calculated[15] by

$$\eta_{\text{max}}^{(\infty-\text{Cascade})} = 1 - \exp\left(-\int_{T_c}^{T_h} \frac{dT}{T} \frac{\sqrt{1+Z_{\text{max}}(T)T}-1}{\sqrt{1+Z_{\text{max}}(T)T}+1}\right) \quad (2)$$

which is strictly increasing with $T_h$. The theoretical maximum efficiency is 25% (≈1/4) at $T_h = 880$ K ($\Delta T = 580$ K). This high efficiency corresponds to a device $ZT$ value of 2. Furthermore, an even higher theoretical maximum efficiency of 33% (≈1/3) is possible at $T_h = 1400$ K ($\Delta T$=1100 K). For single-stage P-N leg devices, the best efficiency increases with $T_h$ and has the highest value of 17.1% (≈1/6) at $T_h = 860$ K ($\Delta T = 560$ K). However, the best efficiency no longer increases but significantly drops when $T_h > 940$ K. This efficiency drop at very high temperatures is due to the absence of a stable material at high temperatures and material self-compatibility issues[15] arising from the strong temperature dependence of thermoelectric properties and low average $ZT$.

A multistage structure in devices such as segmented legs may overcome the self-compatibility issue, as theoretically reported by Ouyang and Li in 2016 ($\eta$ = 21.0% at $\Delta T$ = 700 K)[16]. Higher theoretical efficiencies have been reported in P-type single-leg device: Ryu et al. in 2021 ($\eta$ = 21.9−24.5% at $\Delta T$ = 600−800 K)[12] and Wabi et al. in 2022 ($\eta$ = 22.0% at $\Delta T$ = 600 K)[17]. In addition, in this study, we find a multiple-stage P-N leg pair device with very high efficiency: $\eta$ = 20.0−23.9% at $\Delta T$ = 550–800 K; see Supporting Data file (data700) for the device configuration generated.

**However, it should be noted that there is a large difference between the theoretical and measured best efficiencies; see Fig. 1 and Table 1.** For the single-stage devices in Table 1, the measured best efficiencies range from approximately 7−10%: the single-stage $Bi_2Te_3$-based device by KELK (7.2% at $T_h$ = 553 K)[3] and the GeTe(P)-$Mg_3Sb_2$(N)-based device by Tongji University (10.0% at $T_h$ = 600 K)[18]. For the multistage devices in Table 1, the measured best efficiencies are approximately 12−13%: the GeTe/BiTe(P)-PbTe/$Bi_2Te_3$(N) segmented device by SUSTECH (13.3% at $T_h$ = 873 K)[8], the half-Heusler (HH)/$Bi_2Te_3$ segmented device by PSU (12.0% at $T_h$ = 873 K)[19], the skutterudite (SKD)/BiTe segmented device by SICCAS (12.0% at $T_h$ = 849 K)[20], and the PbTe/$Bi_2Te_3$ cascaded device by AIST (12.0% at $T_h$ = 873 K)[21]. However, their efficiency values are much lower than the theoretically best device efficiency of 17.1% found in this work. Such a loss in efficiency may be due to a suboptimal choice of materials and significant interfacial resistances. Thermal radiation and convection might also cause a nonnegligible loss in efficiency when $T_h$ is high[16].

## Distribution of thermoelectric properties

**Fig. 2** shows the distribution of thermoelectric property values in the thermoelectric big data of **13,338** published material samples. **Fig. 2(a)** shows the available temperature ranges that are defined as the range from the minimum measured temperature to the maximum measured temperature for a given material's thermoelectric property. There are two distinct measurement patterns: (1) cryogenic-type thermoelectric property measurements below room temperature (< 300 K) and (2) power generation-type measurements above room temperature (> 300 K). For the latter, 80% of the samples are measured from 300 K up to 800 K, which are suitable for mid-temperature thermoelectric power generation applications. The highest measured temperature is 1500 K. **Fig. 2(b)** shows the distribution of 215,526 Seebeck coefficient values ($\alpha(T)$). The distribution is somewhat symmetric between the P- and N-type materials. When $\alpha(T)$ is small, it increases with *T*, showing metallic behaviour. When $\alpha(T)$ is large, it increases with *T,* while $T \leq 300$ K, showing bipolar transport behaviour of the narrow-gap semiconductors. **Fig. 2(c)** shows the distribution of 223,404 electric resistivity values. The resistivity varies exponentially with temperature. For temperatures below 300 K, most of the explored samples have very small resistivity, similar to semimetallic or heavily doped semiconductors. Some have very high resistivity, which might be related to charge carrier quenching in doped semiconductors. For temperatures larger than 300 K, the distribution is symmetrical about the axis of the critical value $\rho_{\text{crit}} = 10^{-4}$ Ω m. Furthermore, the resistivity value seems to converge to $\rho_{\text{crit}}$ as the temperature increases. **Fig. 2(d)** shows the distribution of 187,244 thermal conductivity values. When the temperature is below 100 K, there are unusually small and large lattice thermal conductivities. The small value is due to the small number of phonon activation modes, while the large value is due to less phonon scattering and ballistic phonon transport. With increasing temperature, the number of large thermal conductivity samples decreases, which may be due to the activation of the Umklapp three-phonon process. On the whole, the samples can have small thermal conductivities below 10 W m$^{-1}$ K$^{-1}$ above 300 K.

## History of thermoelectric performances: *ZT* and efficiency

For decades, the thermoelectric performance of materials has been developed to search for high *ZT* values. **Fig. 3(a) shows how the best *ZT* value has been improved over time.** In 2000, the *ZT* exceeded 1 for the first time due to nanostructuring and low thermal conductivity[22,23]. Then, the *ZT* finally reached 2 due to the synergetic effect of electron and phonon transport[7,24]; the largest improvement was achieved in the mid-temperature range of approximately 600−950 K. Recent studies report very high peak *ZT* values exceeding 2.5 in GeTe[8] and SnSe[10].

**Fig. 3(b) shows how the best thermoelectric material efficiency ($\eta^{(mat)}$) of single-leg P- or N-type materials has been improved over time for various $T_h$'s and $T_c = 300$ K.** Between 2000 and 2009, the material efficiency was highly enhanced for all temperature ranges, similar to the improvement of *ZT* values. The improvement was the largest at $T_h \approx 900$ K, similar to the improvement of the peak *ZT*. However, after 2010, the increase in the best efficiency for low- and high-temperature applications has been rather small. This is mainly because, even if the peak *ZT* is high, the average *ZT* can be low. Additionally, for wide-temperature applications, the self-compatibility problem[15] of materials may also responsible for the limited efficiency increase. Our observation of the large discrepancy between *ZT* and the material efficiency implies the importance of studying device efficiency.

## Performance by material composition

**Fig. 4 shows the ideal thermoelectric material efficiency over 17 material groups based on material composition; see the Supporting Data file (data261).** In **Fig. 4(a)**, the theoretical material performances are represented for well-known telluride alloys (Te-alloys): $Bi_2Te_3$-, $AgSbTe_2$-, GeTe-, and (Pb,Sn)Te-based alloys. For Te alloys, P-type materials perform better than N-type materials. $Bi_2Te_3$ alloys are the best materials for low-temperature heat sources ($T_h < 600$ K); their best efficiency can reach 10−12%. For mid-temperature heat sources (600 K $< T_h <$ 950 K), P-type $AgSbTe_2$-, GeTe-, and (Pb,Sn)Te-based alloys show superior thermoelectric conversion performance compared to other alloys.

In **Fig. 4(b)**, the thermoelectric material performances of oxide-, sulfide-, and selenide-related alloys (X-O, X-S, X-Se) are provided. Compared to tellurides, these alloys can operate at high temperatures. In the case of $La_2S_3$, although $\Delta T$ is very large (1100 K), its best material efficiency is smaller than that of $Bi_2Te_3$ for $\Delta T = 300$ K. In **Fig. 4(c)**, the performances of Mg and Si alloys are shown. Note that the P-type MgAgSb-based and N-type $Mg_3Sb_2$-based materials have comparable efficiency to $Bi_2Te_3$-based devices for low-temperature heat sources, which explains the experimental device efficiency[25]. For N-type materials, $Mg_2Si$-based alloys exhibit good thermoelectric performance at mid temperatures. Alternatively, $Si_{1-x}Ge_x$ alloys demonstrate good performance at high temperatures ($T_h > 950$ K), as reported[9]. **Fig. 4(d)** shows high-temperature thermoelectric materials such as clathrate (Clath.), HH, and SKD thermoelectric materials and other antimonide compounds (X-Sb). Although their performance is relatively poor at low and mid-temperatures, they show the best efficiency with regard to high-temperature heat sources.

## Representative P-N leg pairs

**Fig. 5 and Table 2** show the theoretical device efficiency curves of 9 representative single-stage P-N leg-pair devices; see **Supporting Data file** (**data600**). Within available temperature ranges, the device efficiencies increase with temperature, indicating that thermoelectric efficiencies are limited by the available temperature ranges, which might be determined by the material thermal stability. For low-temperature heat sources, the fully $Bi_2Te_3$-based P-N leg-pair device shows the best device efficiency, which is higher than 10%. For 600 K < $T_h$ < 950 K, the devices based on P-type chalcogenides and N-type $Mg_3Sb_2$, $Mg_2(Si,Sn)$, HH, or SKD show the highest efficiency. A limit efficiency of 17% is found in the single-stage PbTe-SKD and SnSe-SKD devices. For $T_h$ > 950 K, however, the device efficiency is smaller than that of mid-temperature devices. Notably, there is a distinct pattern in the curves; that is, the curves for low-temperature devices are concave, but the other curves are convex at low temperatures and become linear at high temperatures. The convexity of the curves is related to the poor *ZT* value at low temperatures in addition to the temperature-dependent nature of thermoelectric transport properties. This suggests that the temperature gradient in thermoelectric properties should be controlled using segmented or cascaded device structures. On the other hand, the linearity of the curves at high temperatures implies that linear extrapolation of the curves can be used to estimate the efficiency at higher temperatures.

## Efficiency loss due to interfacial resistances

**Fig. 6** shows the degradation of efficiency of P-N leg-pair devices under electrical and thermal contact resistances compared to the best efficiency of perfect devices; see **Supporting Data file** (**data500**). While the P and N legs generate electrical power, some of the power is lost via internal resistances[16]. The P- and N-type thermoelectric semiconductors are metallized and connected to the electrodes on the substrates, and the device substrates are in contact with the external heat source and sink. Such a complex layered structure causes parasitic electrical and thermal interfacial resistances ($\rho_c$ and $\kappa_c^{-1}$), resulting in net power reduction. The net temperature difference decreases with interfacial thermal resistance, and additional Joule heating occurs at the ends of the legs via additional interfacial electrical resistance. When contact resistances are extremely small ($\rho_c = 10^{-10}$ $\Omega\ m^2$ and $\kappa_c = 10^5$ $W\ m^{-2}\ K^{-1}$), the relative efficiency loss is only 1.5% less than the best device efficiency with perfect contact. For $\rho_c = 10^{-9}$ $\Omega\ m^2$ and $\kappa_c = 10^4$ $W\ m^{-2}\ K^{-1}$, the drop in the best efficiency is approximately 12% on average. For the normal interface condition with $\rho_c = 10^{-8}$ $\Omega\ m^2$ and $\kappa_c = 10^4$ $W\ m^{-2}\ K^{-1}$, the enhanced interfacial Joule heating causes a significant drop in relative efficiency by 22% on average. For an inferior interface condition ($\rho_c = 10^{-7}$ $\Omega\ m^2$ and $\kappa_c = 10^3$ $W\ m^{-2}\ K^{-1}$), a large efficiency loss occurs (67% loss on average), and the best efficiency is highly reduced to 6.5% (≈1/15).

## Summary

The achievable **best thermoelectric device efficiencies** are theoretically investigated **over ever-explored materials**. Theoretically, an efficiency of 1/3 is possible in an infinite-cascade device. An efficiency of 1/4 is possible in a segmented device, while an efficiency of 1/6 is possible in a single-stage device. However, these theoretical limits are much higher than the measured device efficiencies of 1/8 and 1/10 of the multistage and single-stage devices. A poor interface quality may yield a low conversion efficiency of 1/15 to 1/14. The discrepancy between theoretical and experimental efficiency can be mitigated in multistage devices by reducing interfacial resistance and selecting optimal thermoelectric materials.

## Methods

**Overview of the entire process to find the best efficiency.** The best thermoelectric efficiency is calculated and explored using the following procedure. The values in parentheses at the end of the following sentences are the corresponding identification codes for the analysis and data. First, a formatted thermoelectric property file (**data010**) is generated by extracting the raw data from Starrydata2. A sample metadata file (**data030**) is also generated; it contains thermoelectric data length, data reference, composition information, available temperature range, published date, etc. A data filter is used to test that each material sample's dataset of thermoelectric properties is complete, valid, or errorless, and the result is recorded as sample lists (**data040**). The thermoelectric properties are interpolated (**data070**) for the $ZT$ distribution analysis. Composition data are transformed into a composition vector, and the samples are classified into given material groups (**data150**). The thermoelectric material efficiency of the samples in the list is computed for available temperature ranges (**data234**), and the efficiency data are interpolated (**data261**). Using high-efficiency samples, P-N leg pair device configurations are generated (**data300**), and the device efficiencies are computed (**data400**). Finally, the best $ZT$, best material efficiency, and best device efficiency curves are obtained (**data500**). The best efficiencies are compared to the experimentally reported device efficiencies (**data500**). Among them, 9 representative best efficiency devices are analysed (**data600**). A multiple-stage P-N leg pair device with a very high efficiency of 23.9% is found (**data700**).

**Data preparation, filtering, and cleansing.** High-quality thermoelectric property data are obtained from the **Starrydata2** thermoelectric web-DB[13], which is growing with time. For the investigations, we use the thermoelectric property data of version 2021-November-24, consisting of 43,601 material samples from 7,994 published papers ("20211124.zip" in https://github.com/starrydata/starrydata_datasets). Among them, 16,420 samples are chosen from 3,585 published papers that contain a full set of three thermoelectric properties. Note that a complete set of thermoelectric properties is mandatory for the calculation of thermoelectric performance. Then, high-quality thermoelectric property data are obtained after filtering out unphysical and erroneous data. A total of 33 data filters are used, that is, *physics filters* to remove unphysical values of thermoelectric properties and temperature ranges and *error filters* to remove insufficiently/incorrectly labelled data that are hardly readable by a computer. See Supporting Data file (**data040**) for the thermoelectric sample filtering table. For high-efficiency samples, the correctness of the thermoelectric properties from Starrydata2 is confirmed by visual inspection. As a consequence, high-quality thermoelectric property data are obtained, which consist of 13,338 complete and valid samples from 3,121 published papers. The corresponding data size is 215,526 $(T, \alpha(T))$ pairs, 223,404 $(T, \rho(T))$ pairs, and 187,244 $(T, \kappa(T))$ pairs. For efficiency calculations, thermoelectric properties are interpolated in a piecewise linear manner and extrapolated in a constant value manner so that the resulting properties are continuous functions of temperature. Among the complete and valid samples, 12,645 samples have thermoelectric properties for temperatures above 300 K. Using these samples, thermoelectric efficiencies at $T_h > 300$ K are evaluated for various device models.

**Thermoelectric device model.** In the thermoelectric device, P- and/or N-type thermoelectric materials, called thermoelectric legs, are placed between hot- and cold-side substrates. The thermal boundary conditions are assumed to be the Dirichlet condition, which means that $T_h$ and $T_c$ are fixed during device operation. We compute the theoretical maximum conversion efficiency, ignoring radiation and convection losses. The length of a thermoelectric leg is assumed to be 3 mm. Since the leg is connected to substrates, we consider the interfacial electrical and thermal resistances. The electrical and thermal resistances are simultaneously imposed using two additional segments of 0.1 mm attached at the ends of the leg unless interfacial resistances are zero; in total, the leg length is 3 mm for a perfect interface and 3.2 mm if there are interfacial resistances. We assume that the electrical and thermal currents flow perpendicular to the substrate, which implies a one-dimensional flow. For a single-leg device, the leg cross-sectional area of $A = $ 3 mm × 3 mm is adopted. For a P-N leg-pair device, the cross-sectional areas of P- and N-type legs ($A_p, A_n$) are set to $A_p = (1-x) \cdot A$ and $A_n = x \cdot A$, for a number $x$ between 0.02 and 0.98. For $x$, we consider 11 values at the Chebyshev nodes of the second kind between 0.02 and 0.98.

**Thermoelectric performance in a one-dimensional leg.** In a one-dimensional thermoelectric leg, the heat currents at the hot and cold sides ($Q_h, Q_c$) and the power $P$ are given as follows:

$$Q_{h,c} = -A\kappa_{h,c} \left(\frac{dT}{dx}\right)_{h,c} + I\,\alpha(T_{h,c})T_{h,c}\,, \qquad (3)$$

$$P = Q_h - Q_c\,, \qquad (4)$$

where $h$ and $c$ denote the hot and cold sides, $A$ is the leg cross-sectional area, and $I$ is the electric current flowing through the leg. Once the temperature distribution inside the leg is known, the thermoelectric material efficiency can be easily computed for given electrical and thermal conditions:

$$\eta^{(\mathrm{mat})} = \eta(I, T_h, T_c) = \frac{Q_h - Q_c}{Q_h}. \tag{5}$$

The P-N leg-pair device efficiency can be computed as follows:

$$\eta^{(\mathrm{dev})} = \frac{P_{tot}}{Q_{tot}} = \frac{P^{(p)} + P^{(n)}}{Q_h^{(p)} + Q_h^{(n)}}. \tag{6}$$

**Temperature distribution inside a one-dimensional leg.** In a steady-state one-dimensional leg, thermoelectric effects are governed by the following thermoelectric differential equations[12,26]:

$$\frac{d}{dx}\left(\kappa \frac{dT}{dx}\right) - T\left(\frac{d\alpha}{dT}\right)\left(\frac{dT}{dx}\right)J + \rho J^2 = 0. \tag{7}$$

This equation can be transformed into a thermoelectric integral equation for temperature distribution $T(x)$ via double integration on $f_T(x) \coloneqq -T\frac{d\alpha}{dT}\frac{dT}{dx}J + \rho J^2$ as follows[12]:

$$T(x) = \left(T_h - \frac{K\Delta T}{A}\int_0^x \frac{1}{\kappa(s)}ds\right) \\ + \left(-\int_0^x \frac{F_T(s)}{\kappa(s)}ds + \frac{K\delta T}{A}\int_0^x \frac{1}{\kappa(s)}ds\right) \tag{8}$$

where $F_T(x) \coloneqq \int_0^x f_T(s)ds$, $\delta T \coloneqq \int_0^L \frac{F_T(x)}{\kappa(x)}dx$, and $L$ is the leg length. By iteratively computing **(8)**, we find the temperature distribution inside the legs[12].

**Computation of the best efficiency.** A thermoelectric sample is evaluated using the maximum thermoelectric efficiency calculation under various temperature differences $\Delta T = T_h - T_c$, where the cold-side temperature is 300 K and $T_h$ is chosen between 301 K and the maximum available temperature ($T_{\max}$) plus 15 K. For $T_h$, we consider 11 values at the Chebyshev nodes of the second kind between 301 K and $T_{\max}$.

The generated electrical power and input heat current are computed using **equations (3-8)**[12] under a given electrical current and thermal boundary conditions. Then, the maximum efficiencies are searched by varying electrical currents. Using the searched high-performance P- and N-samples, of which the efficiency is larger than or equal to 85% of the best efficiency, single-stage P-N leg-pair configurations of 14,702 are constructed; see Supporting Data file (**data300**). We also consider 11 leg-pair geometries and 5 interfacial resistance conditions. In total, 63,225(= 12,645 × 5) single-leg devices and 808,610(=14,702× 11 × 5) P-N leg-pair devices are considered. For each device configuration, 11 electrical current points and 11 thermal boundary conditions are considered. During this process, the leg geometry parameter $x$, current $I$, and temperature $T_h$ are sampled at the Chebyshev node of the second kind, which is suited for polynomial interpolation[27]. In total, we calculate $\eta^{(\text{mat})}$'s for 7,650,225 configurations over (material, interface, $I$, $T_h$) and $\eta^{(\text{mat})}$'s for 97,841,810 configurations over (material leg pair, geometry $x$, interface, $I$, $T_h$). Finally, we explore the thermoelectric material and device efficiency space and obtain the best thermoelectric efficiencies.

**Classification of compositions according to material group.** Samples are grouped into 17 material groups based on composition analysis. We manually define the 17 material groups with representative compositions. To classify a sample into a material group, the fraction of host or anionic elements is calculated. If the fraction is greater than a certain value (90% for the host, 30% for anionic analysis, 20% for oxide classification), then the sample is grouped into one of the 17 material groups of which the representative composition is most similar to the sample's composition. For example, $(Bi_{0.4}Sb_{1.6})Te_3Ag_{0.03}$ is similar to $Bi_2Te_3Ag_{0.03}$ and is grouped into $Bi_2Te_3$. The classification results can be found in the Supporting Data file (**data150**).

## Data availability

Thermoelectric property data used during the current study is available in the GitHub repository (https://github.com/starrydata/starrydata_datasets, raw-data version 20211124.zip) and in the Starrydata2 open-web DB (https://www.starrydata2.org/). Any additional information required to reanalyse the data reported in this paper is available from the corresponding author upon request. **Supporting Data files can be made public after publication.**

## Acknowledgements

This work was supported by the Korea Electrotechnology Research Institute (KERI) Primary research program through the National Research Council of Science & Technology (NST) funded by the Ministry of Science and ICT (MSIT) (23A01002, Research on high-power low-mid temperature thermoelectric power generator via thermoelectric data manifold exploration and expedition), by the Korea Institute of Energy Technology Evaluation and Planning (KETEP) grant funded by the Ministry of Trade, Industry and Energy (MOTIE) (2021202080023D, Development of thermoelectric power device-system performance evaluation method and thermoelectric performance prediction technology), and by the National Research Foundation of Korea (NRF) grant funded by the Korea government (MSIT) (MAIN-SUB: 2022M3C1C8093916-2022M3C1C8097621), the Republic of Korea. Y.K. and M.K. were supported by the Japan Science and Technology Agency (JST) CREST grant number (JPMJCR19J1, Development of innovative functional materials based on large-scale search for new crystals), Japan.

## Author contributions

B.R. conceptualized the study, led the design of the work, filtered the data, performed the efficiency calculations, analysed the efficiency data, created the figures, and wrote the manuscript. J.C. developed the software, analysed the efficiency data, discussed the results, and wrote the manuscript. S.P. supervised the project and discussed the results. M.K. and T.M. developed the Starrydata web system and generated the dataset file. Y.A., S.G., A.T., D.Y. and M.F. collected thermoelectric property data from publications. S.G., Y.I., A.T. and Y.K. collected target publications. M.K. and Y.K. served as an advisor on the project and discussed the results.

## Competing interests

The authors declare no competing interests.

## Additional information

**Supporting Data files** are available. Correspondence and requests for materials should be addressed to B.R.

# Tables

**Table 1. Selected values of the best thermoelectric efficiency ($\eta$).** The efficiencies of multistage and single-stage thermoelectric devices are described with working temperatures, Carnot efficiency ($\eta_{\text{Carnot}} = \Delta T/T_h$), and reduced thermoelectric efficiency ($\eta_{\text{TE}} = \eta_{\text{Carnot}}/\eta$).

|   | Device | $T_h$ [K] | $T_c$ [K] | $\Delta T$ [K] | $\eta$ | $\eta_{\text{Carnot}}$ | $\eta_{\text{TE}}$ | Ref. |
|---|---|---|---|---|---|---|---|---|
| Calc. | Infinite cascade | 1400 | 300 | 1100 | 33.0% ≈1/3 | 78.6% | 42.0% | This work |
| | Segmented (P-N) | 1100 | 300 | 800 | 23.9% ≈1/4 | 72.7% | 32.8% | This work |
| | Single-stage (P-N) | 860 | 300 | 560 | 17.1% ≈1/6 | 65.1% | 26.2% | This work |
| Expt. | Segmented (P-N) | 800 | 294 | 506 | 13.3% ≈1/8 | 63.3% | 21.0% | Ref.[8] |
| | Single-stage (P-N) | 600 | 280 | 320 | 10.0% ≈1/10 | 53.3% | 18.8% | Ref.[18] |
| | | 553 | 303 | 250 | 7.2% ≈1/14 | 45.2% | 15.9% | Ref.[3] |

**Table 2. Thermoelectric device efficiencies of 9 representative thermoelectric P-N leg-pair devices and corresponding P- and N-leg compositions and sample ID (sampleid) numbers in Starrydata2.**

| P-leg composition (Ref., sample ID) | N-leg composition | $T_h$ [K] | $\Delta T$ [K] | $\eta^{(\text{dev})}$ | $A_p/A_n$ |
|---|---|---|---|---|---|
| $Si_{0.8}Ge_{0.2}$ /P (Ref.[28], sampleid=21962) | $Si_{80}Ge_{20}$ (Ref.[29], sampleid=21190) | 1285 | 985 | 13.02% | 0.939 |
| $(Nb_{0.60}Ta_{0.40})_{0.8}Ti_{0.2}FeSb$ (Ref.[30], sampleid=31566) | $Si_{80}Ge_{20}$ (Ref.[31], sampleid=21211) | 1185 | 885 | 16.15% | 0.426 |
| $Pb_{0.9}Na_{0.02}Mg_{0.08}Te$ (Ref.[32], sampleid=300) | $(Hf_{0.6}Zr_{0.4})NiSn_{0.99}Sb_{0.01} - W_{0.087}$ (Ref.[33], sampleid=38585) | 940 | 640 | 16.70% | 3.615 |
| $Sn_{0.97}Na_{0.03}Se_{0.9}S_{0.1}$ (Ref.[34], sampleid=41016) | $Yb_{0.3}Co_4Sb_{14.4}$ (Ref.[35], sampleid=41697) | 885 | 585 | 17.05% | 3.181 |
| $Na_{0.035}Eu_{0.03}Mn_{0.03}Pb_{0.905}Te$ (Ref.[36], sampleid=35891) | $xCo/Ba_{0.3}In_{0.3}Co_4Sb_{12}$, x=0.2% (Ref.[37], sampleid=31358) | 860 | 560 | 17.01% | 3.391 |
| $Ge_{0.89}Sb_{0.1}In_{0.01}Te$ (Ref.[38], sampleid=31973) | $Mg_2(Si_{0.4}Sn_{0.6})Sb_{0.018}$ (Ref.[39], sampleid=9777) | 785 | 485 | 16.45% | 1.912 |
| $AgSbTe_2$ (Ref.[40], sampleid=16668) | $Mg_{3.15}Mn_{0.05}Sb_{1.5}Bi_{0.49}Se_{0.01}$ (Ref.[41], sampleid=27133) | 655 | 355 | 13.68% | 0.961 |
| $Bi_{0.4}Sb_{1.6}Te_3Ag_{0.003}$ (Ref.[42], sampleid=38722) | $Bi_{0.24}Sb_{0.05}Te_{0.61}Se_{0.10}$ (Ref.[43], sampleid=38264) | 585 | 285 | 11.67% | 1.106 |
| $Bi_{0.5}Sb_{1.5}Te_3$ (Ref.[6], sampleid=42662) | $Y_{0.2}Bi_{1.8}Se_{0.3}Te_{2.7}$ (Ref.[44], sampleid=16900) | 435 | 135 | 7.75% | 1.034 |

BR et al., arXiv:2210.08837  https://arxiv.org/abs/2210.08837

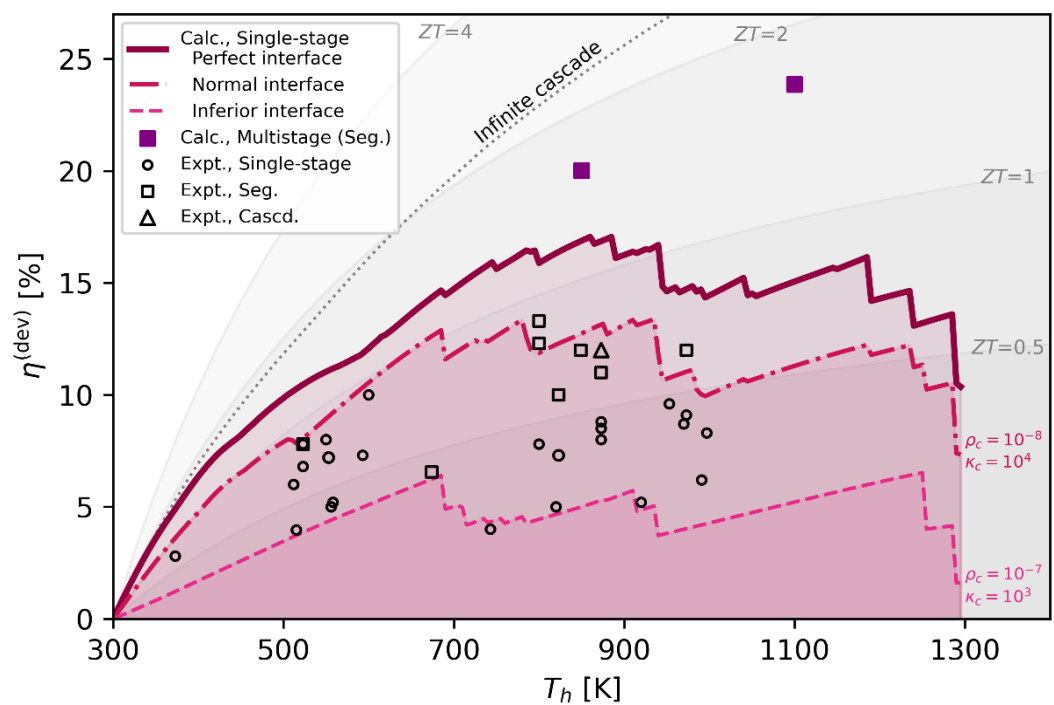

**Fig. 1 Theoretically best device efficiency ($\eta^{(\mathrm{PN})}$) among 808,610 single-stage P-N leg-pair thermoelectric devices when $T_c = 300$ K.** The dotted line is the best efficiency curve for an infinite-cascade device. Solid, dot-dashed, and dashed lines are the best efficiency curves for single-stage devices with a perfect interface (no thermal and electrical resistance), a normal interface ($\rho_c^{(\mathrm{normal})} = 10^{-8}$ Ω m$^{-2}$ and $\kappa_c^{(\mathrm{normal})} = 10^{4}$ W m$^{-2}$ K$^{-1}$), and an inferior interface ($\rho_c^{(\mathrm{inferior})} = 10^{-7}$ Ω m$^{-2}$ and $\kappa_c^{(\mathrm{inferior})} = 10^{3}$ W m$^{-2}$ K$^{-1}$), respectively. Filled squares are efficiencies for calculated (Calc.) multistage segmented P-N leg-pair devices working at $T_h = 850$ K and $T_h = 1100$ K. Additional data points represent the experimental (Expt.) efficiencies of fabricated devices: single-stage devices (unfilled black circle)[3,21,25,43,45–60], segmented devices (Seg., unfilled black square)[8,19,20,45,52,61–63], and a cascaded device (Cascd., unfilled black triangle)[21]. The **ZT** values of the grey-guide lines are inversely calculated using the maximum efficiency equation (1) for a given $T_h$ when $T_c = 300$ K[64]. See Supporting Data file for the best efficiency curves.

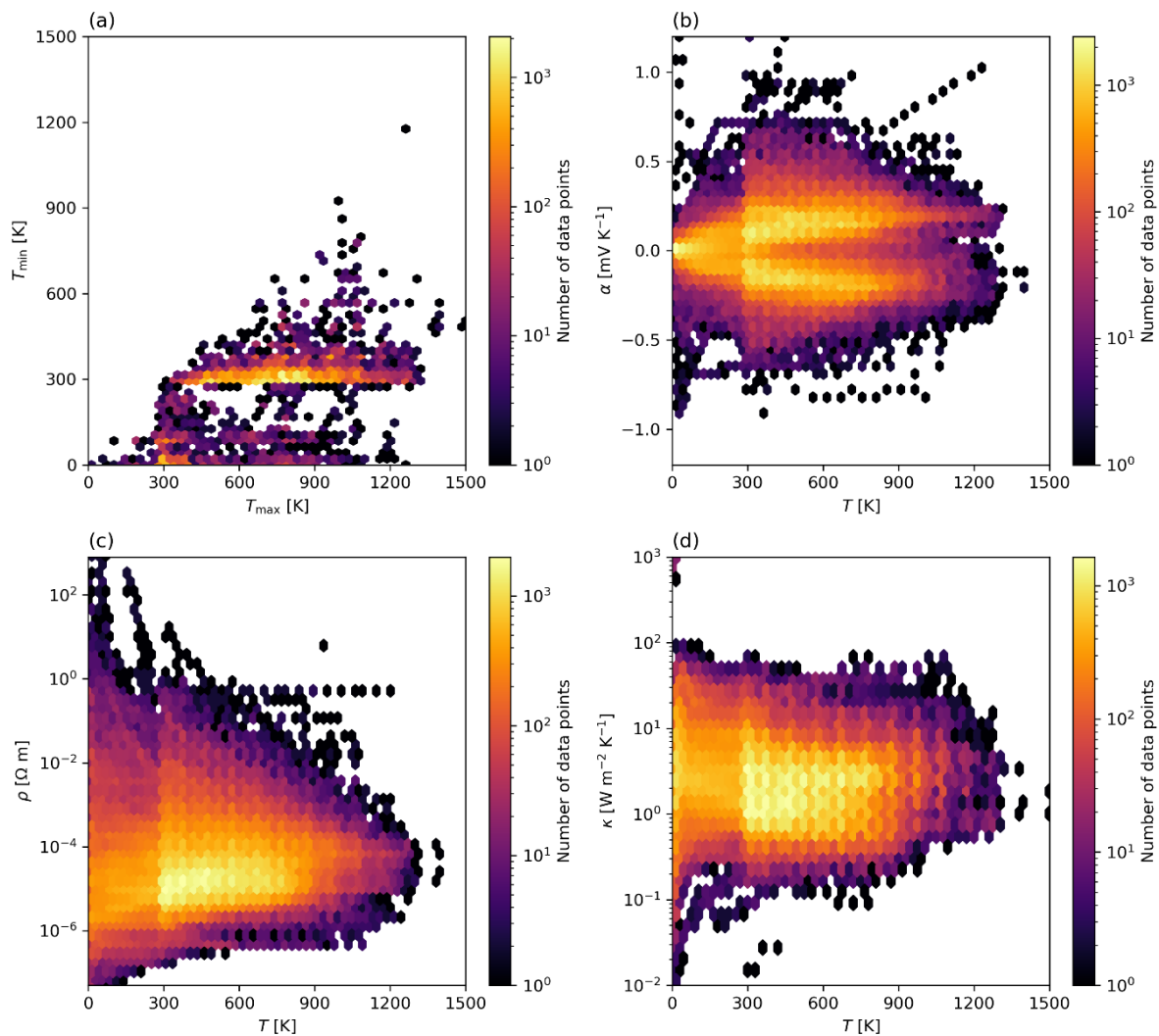

**Fig. 2 Distribution of thermoelectric properties from 13,338 published samples.** (a) Hexa-bin plot of the maximum and minimum available temperatures from the samples' thermoelectric properties. Hexa-bin plots of the (b) Seebeck coefficient distribution, (c) electrical resistivity distribution, and (d) thermal conductivity distribution over measured temperatures. In each panel, colour represents the number of data points.

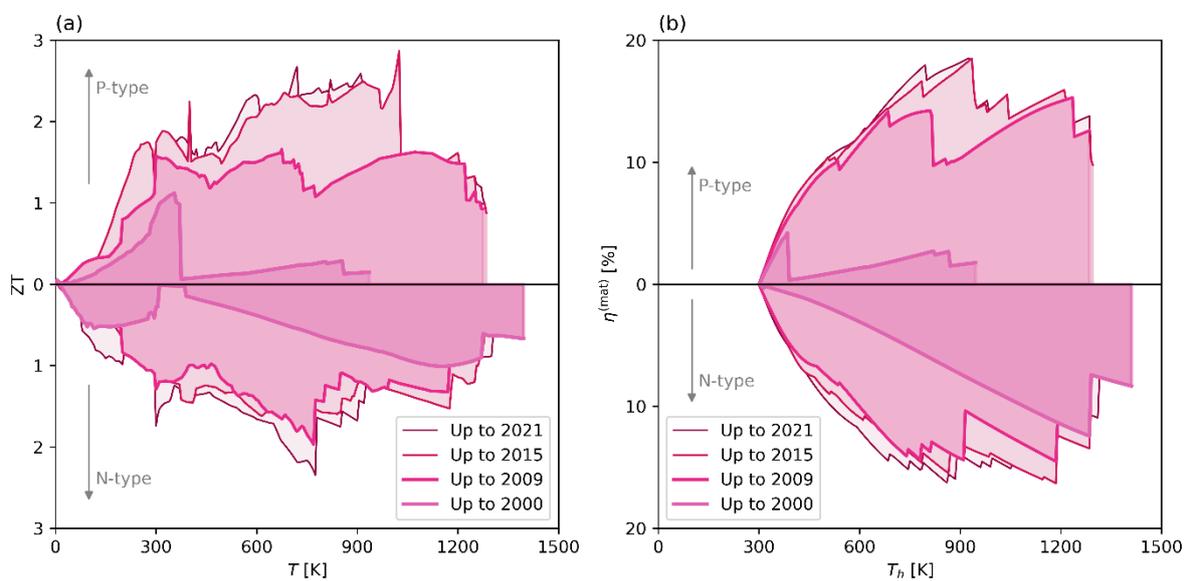

**Fig. 3 Advances in thermoelectric materials' performance.** (a) The best material figure of merit $ZT$ and (b) best thermoelectric material efficiency $\eta^{(\mathrm{mat})}$ for the given year.

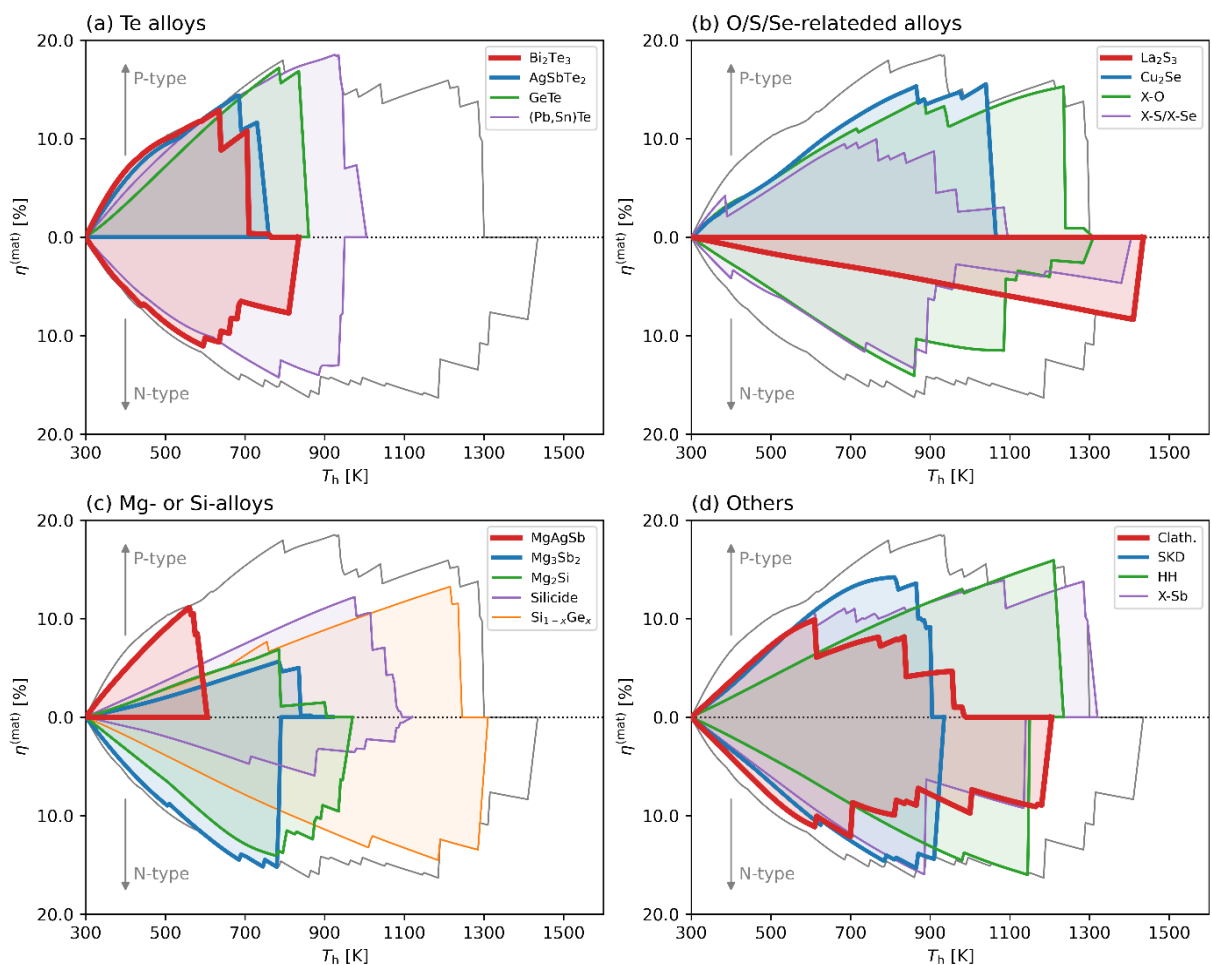

**Fig. 4 Achievable thermoelectric material efficiency for 17 material groups.** (a) Te alloys, (b) O/S/Se-related alloys, (c) Mg- or Si-based alloys, and (d) other alloys. See **Supporting Data files (data261)** for sample material group classification.

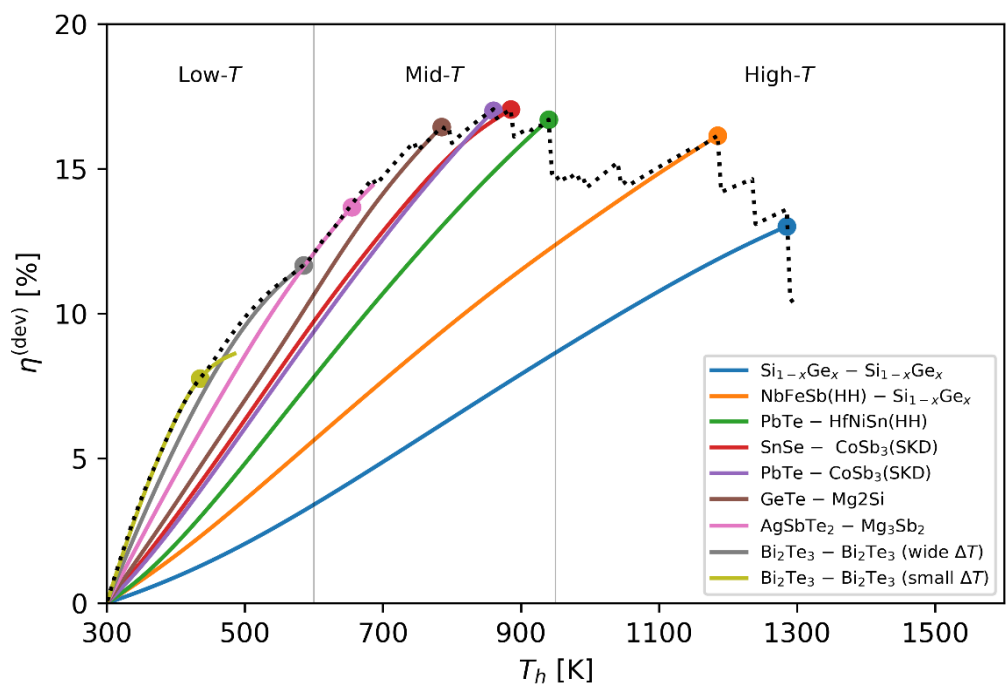

**Fig. 5 Thermoelectric device efficiency curves of 9 representative devices with respect to various heat-source temperatures ($T_c$ = 300 K).** Low-$T$ (<600 K), mid-$T$ (600 K−950 K), and high-$T$ (>950 K) range applications are separated with vertical grey lines. The dotted line is the achievable best device efficiencies using single-stage P-N leg devices. The dot on each curve means that the device has realized the achievable best efficiency. The best efficiency is computed with leg-area ratio optimization. See **Table 2** and the Supporting Data file (**data415**) for detailed information on the designed P-N leg-pair devices.

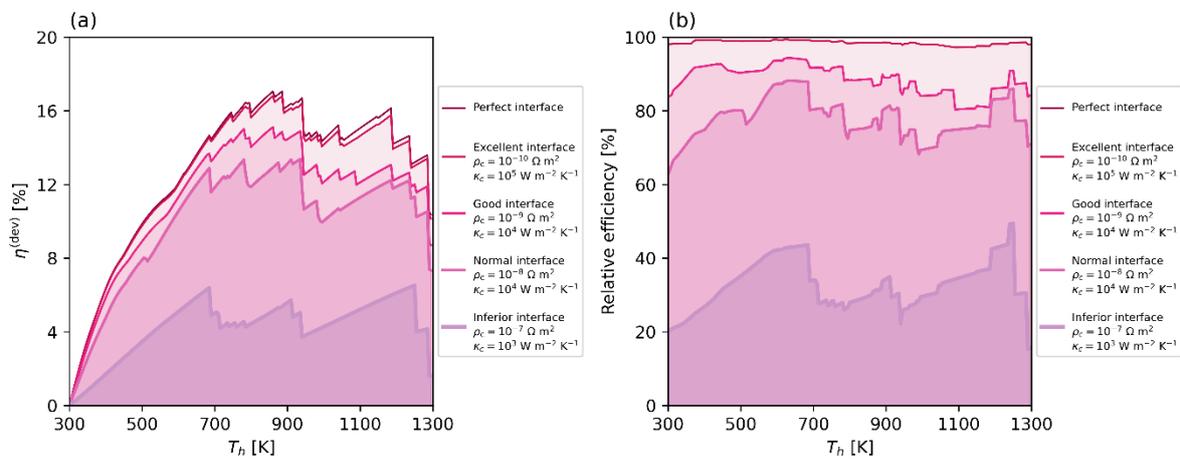

**Fig. 6 Thermoelectric device efficiency under interfacial electrical and thermal resistances.** (a) Achievable device efficiency under interfacial electrical and thermal resistances. (b) Relative efficiency of the best device efficiency under interfacial electrical and thermal resistances compared to the perfect interface devices. See Supporting Data file (**data500**) for detailed information.

# Vector Images: Figs. 1-6 in

**BR et al., "Best Thermoelectric Efficiency of Ever-Explored Materials", arXiv:2210.08837**

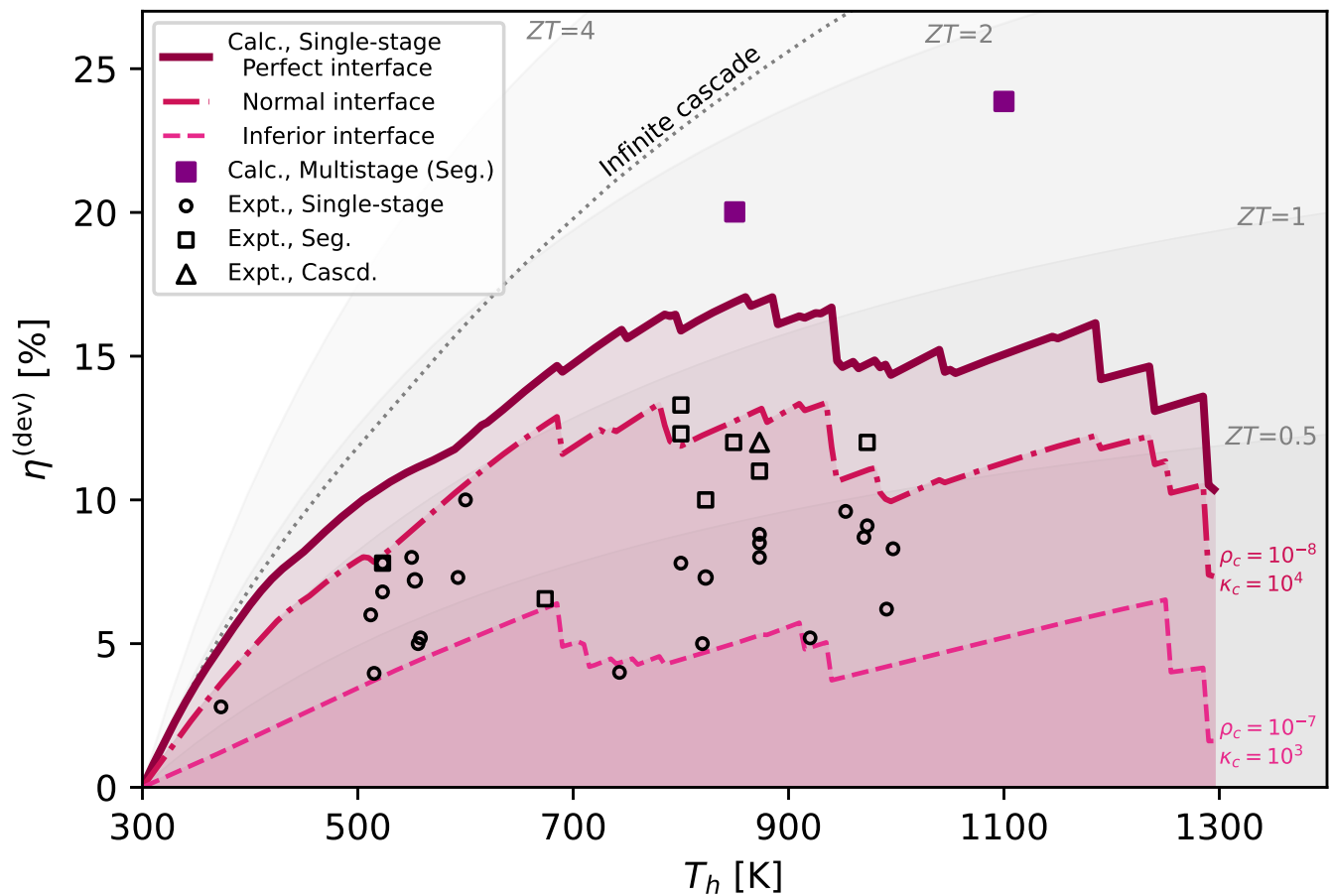

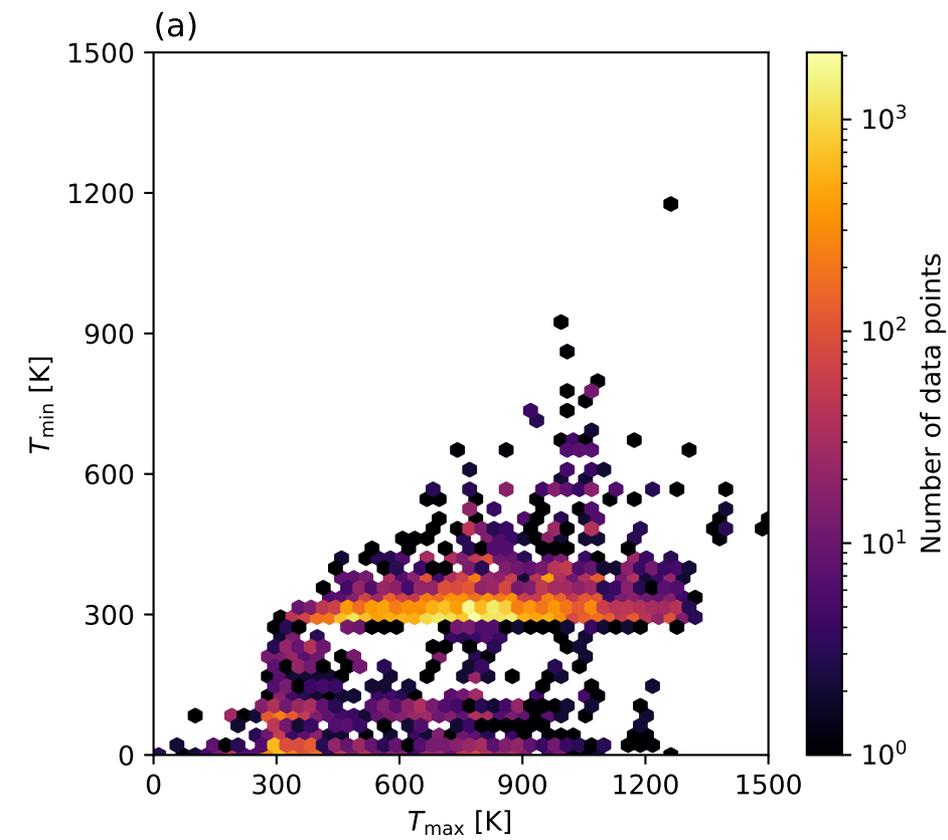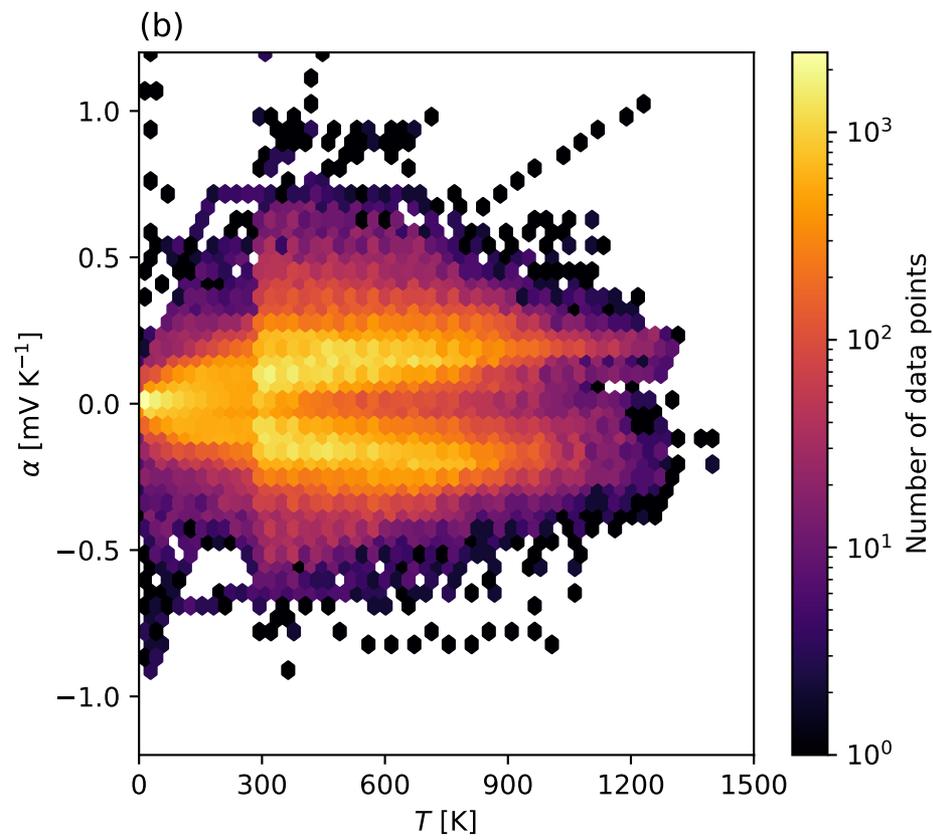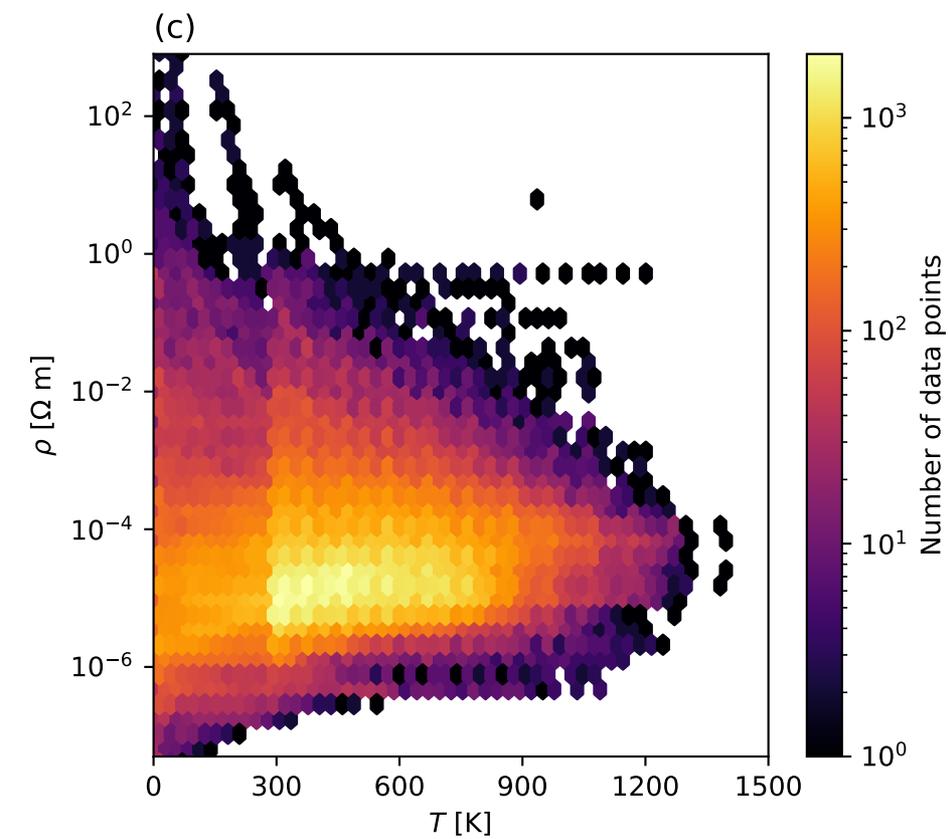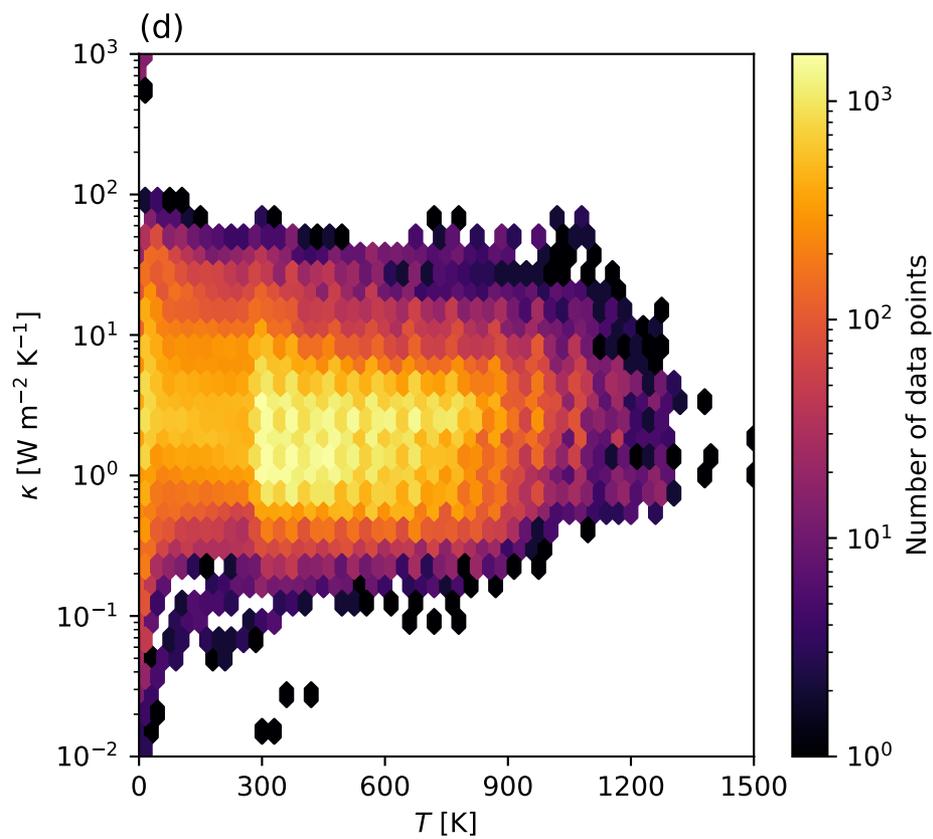

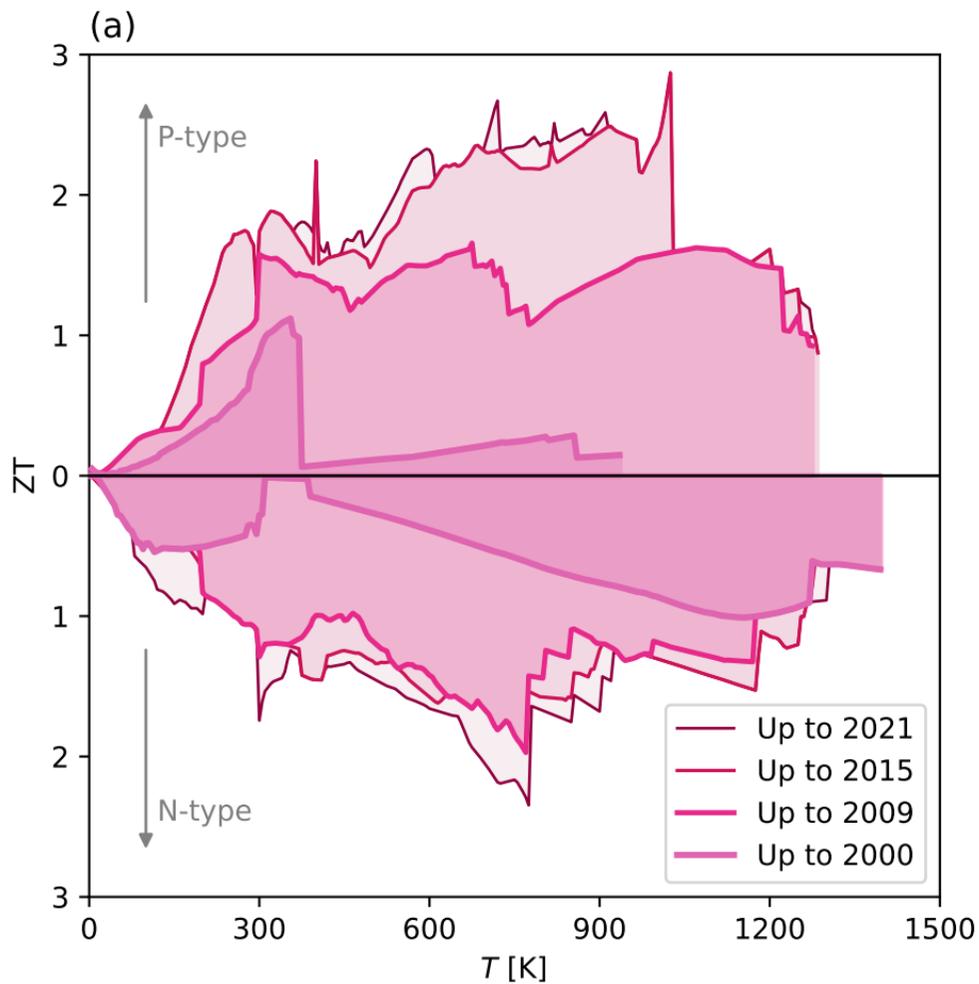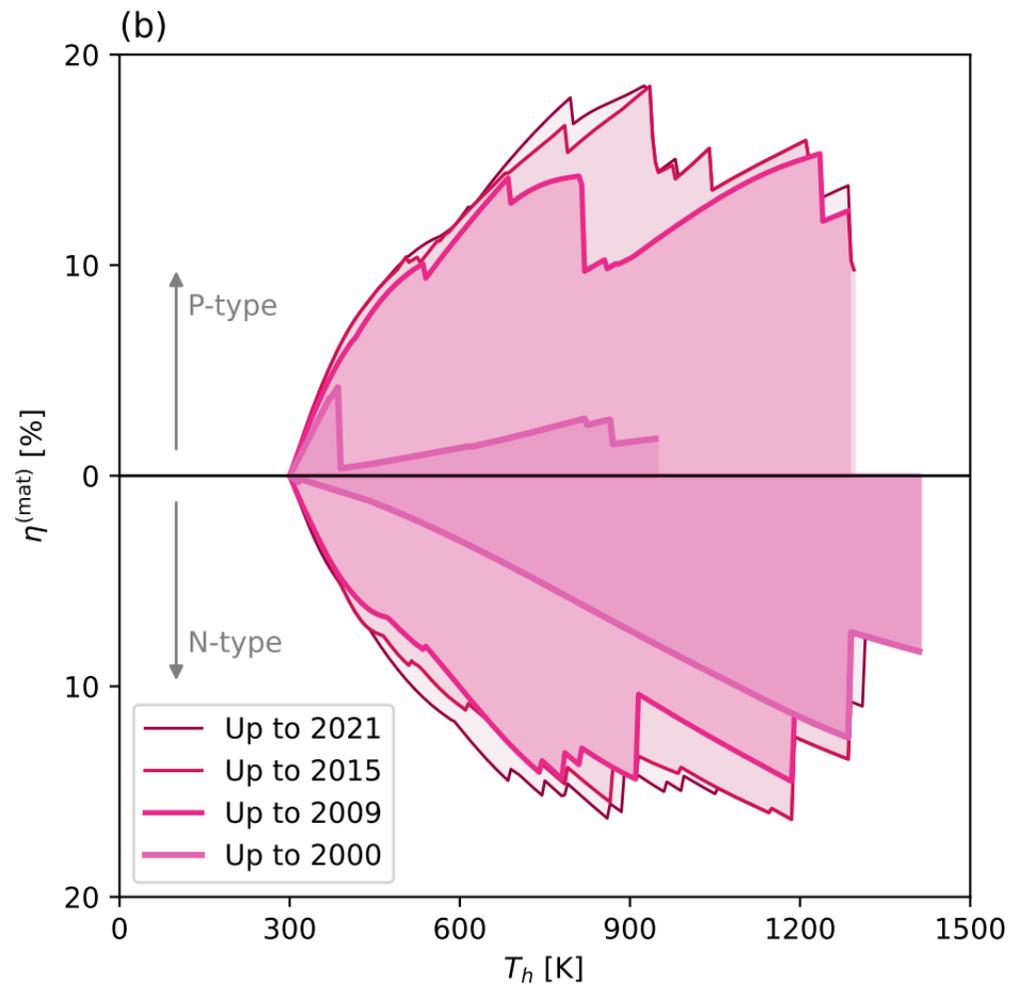

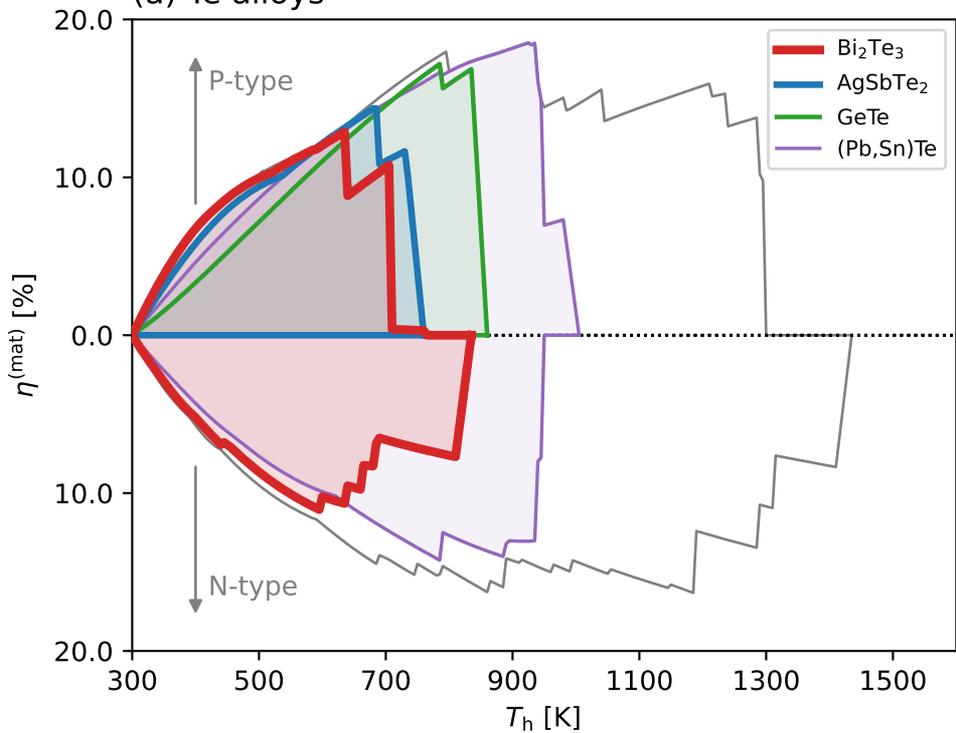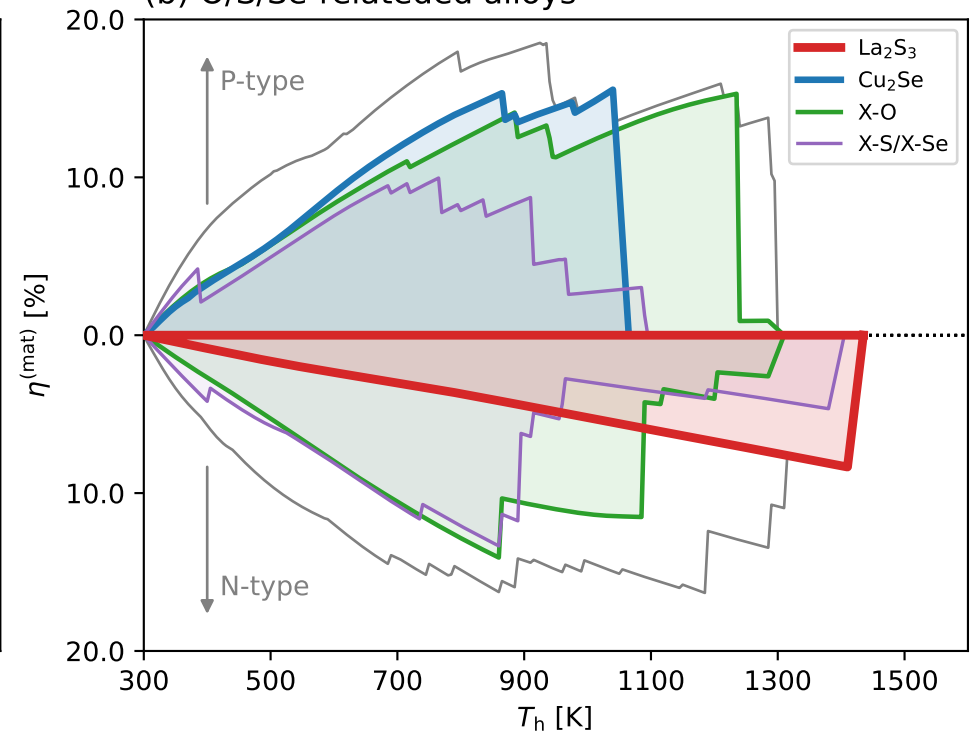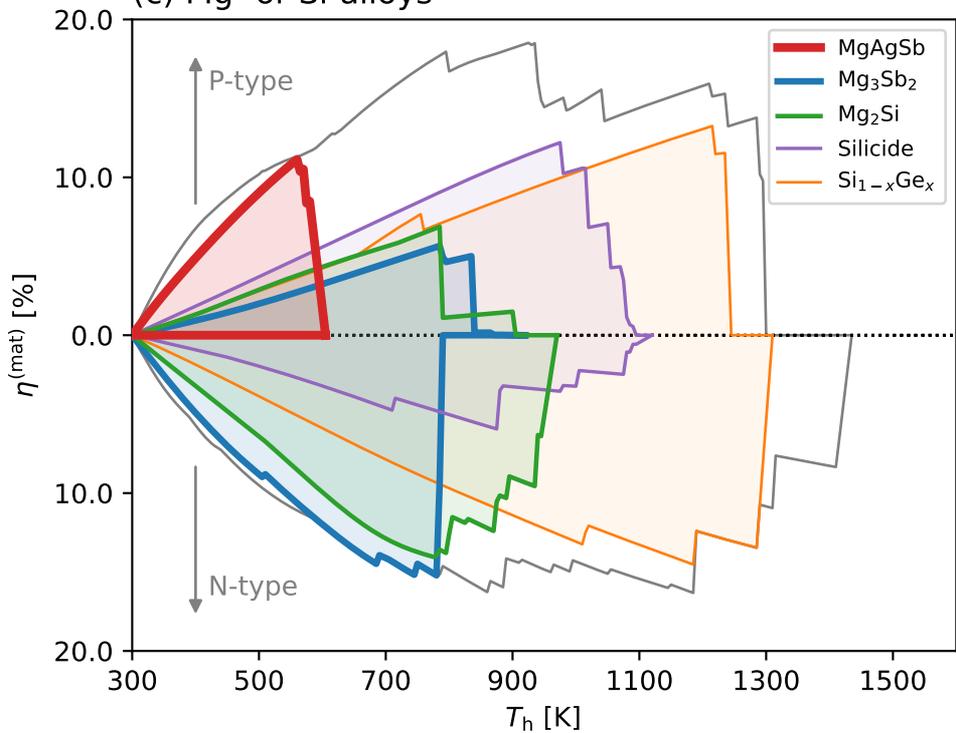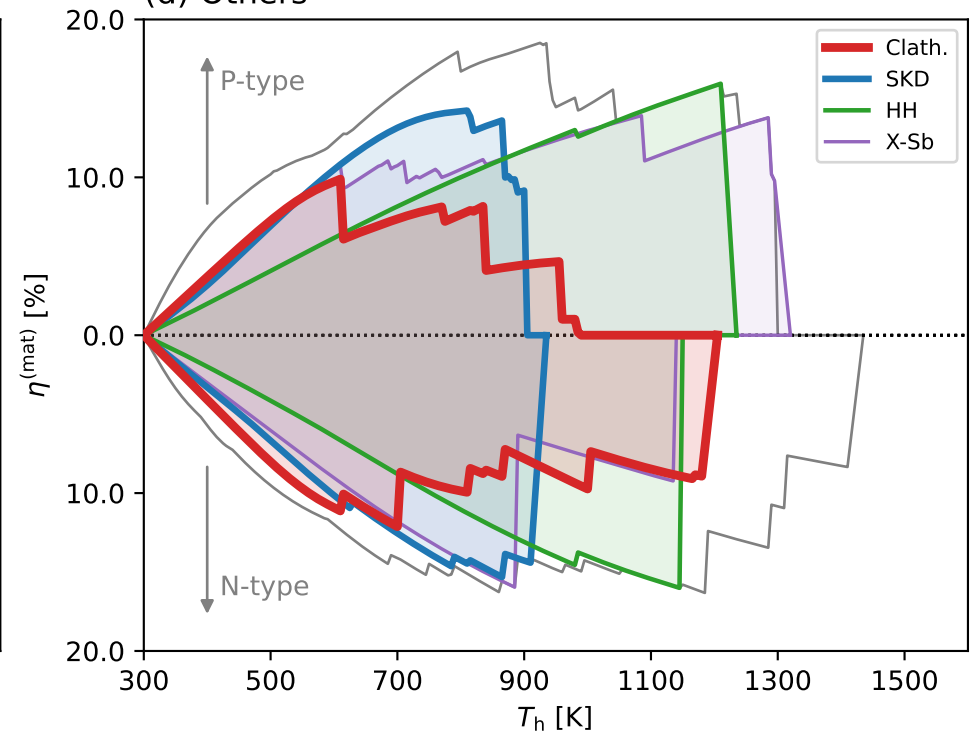

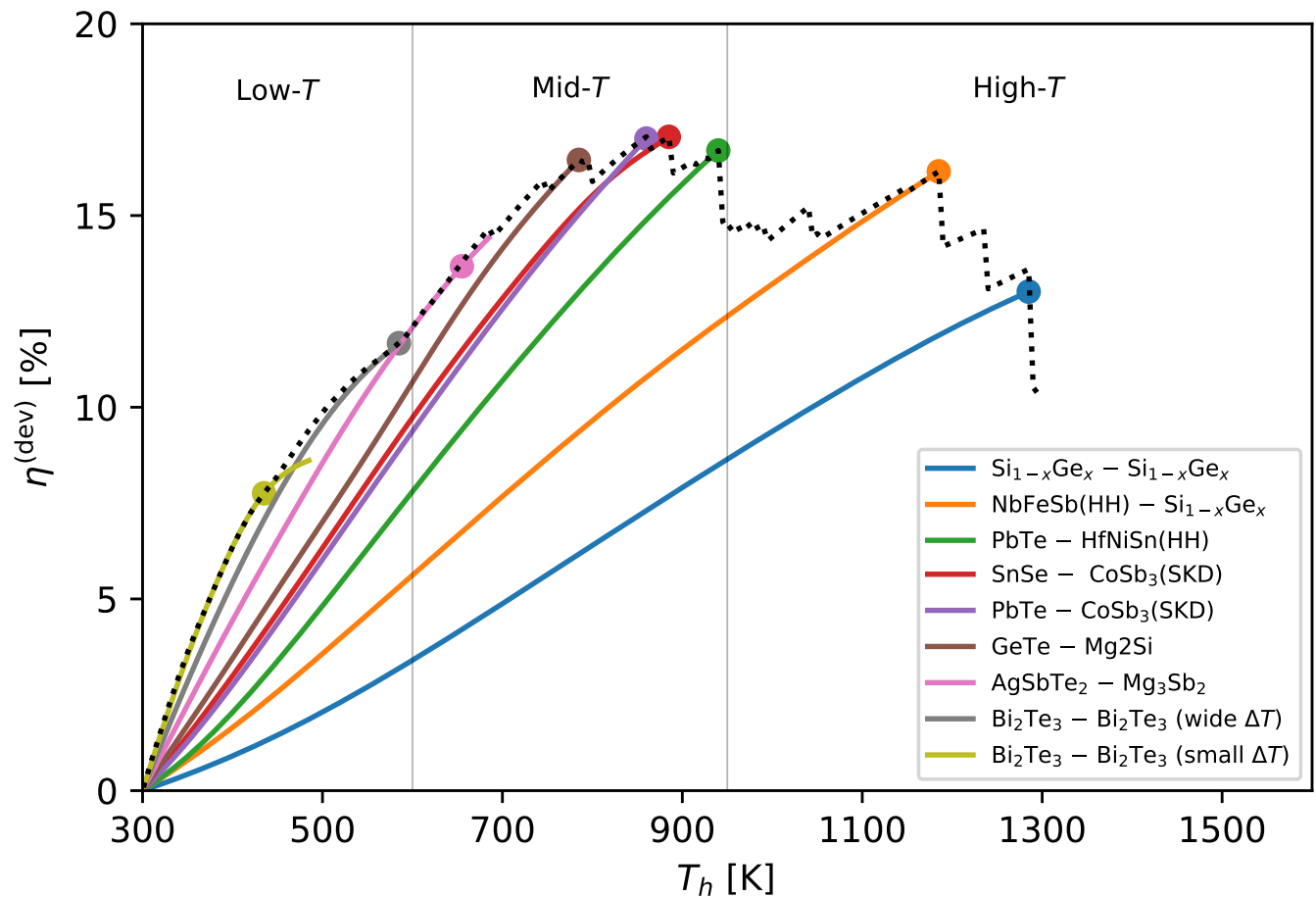

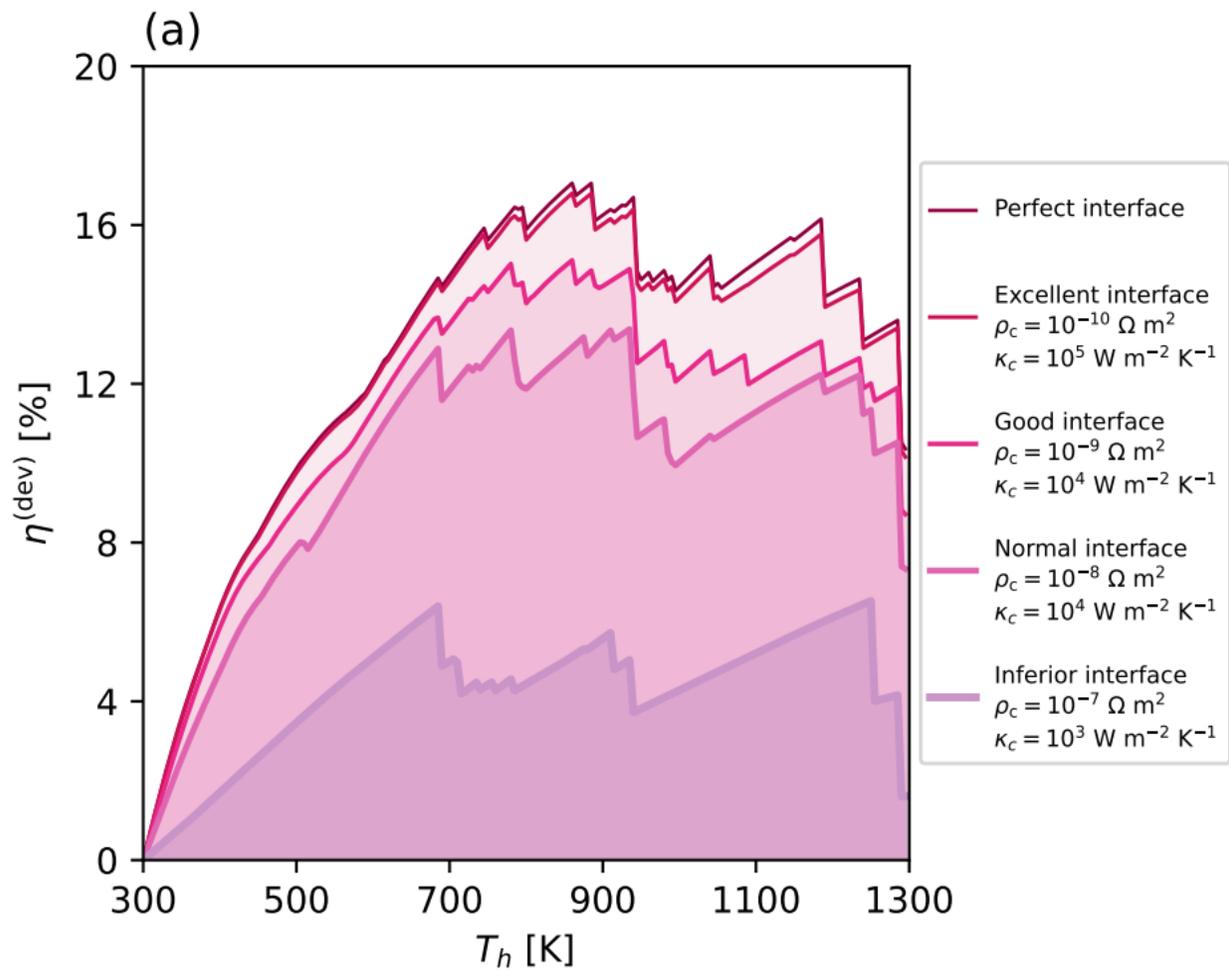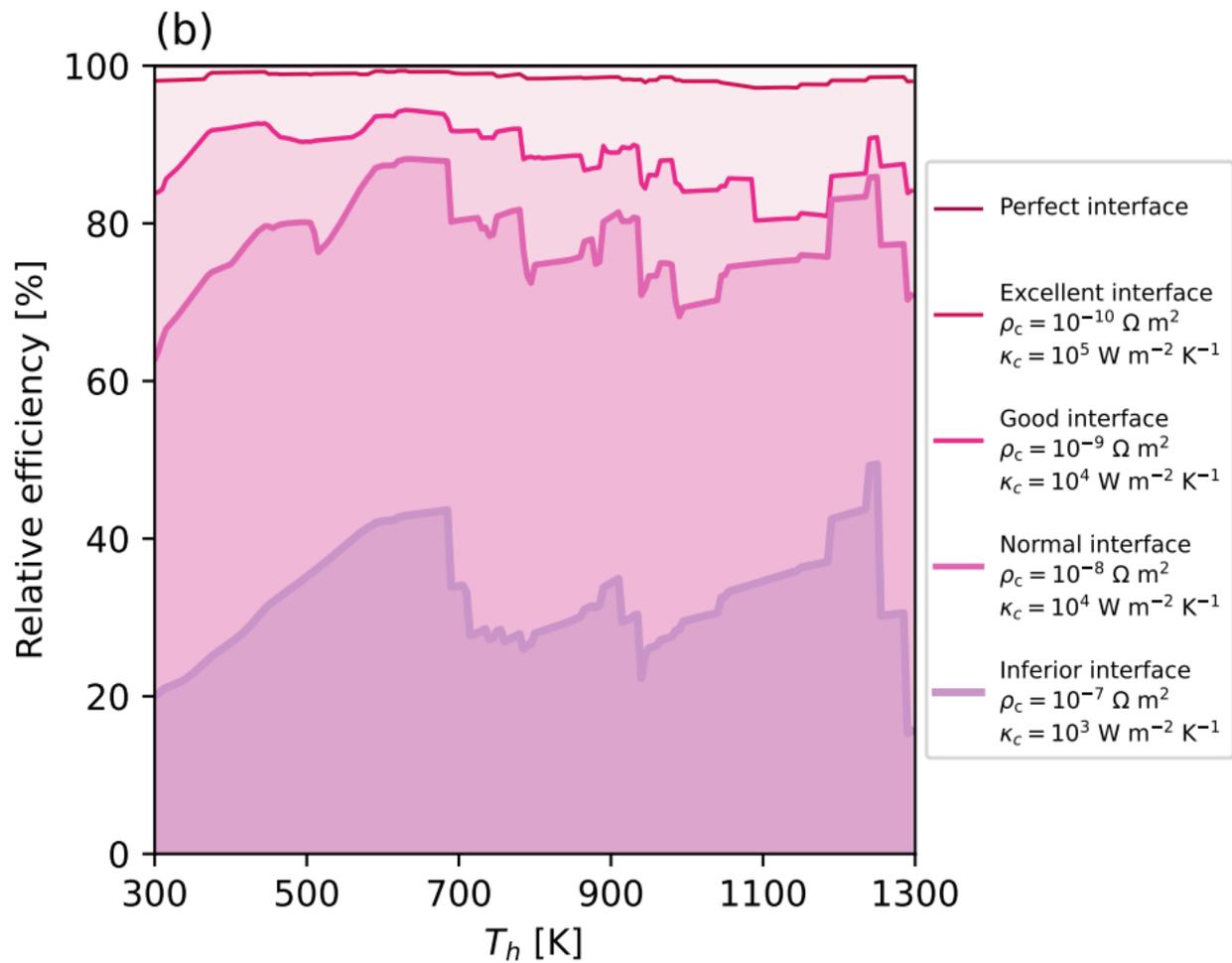